\documentstyle[12pt]{article}
\input epsf
\newcommand{\svec}[1]{\mbox{\boldmath $#1$}}
\begin{document}
\title{\tt An aberrant ``spin-orbit interaction'' persists
in the literature since more than thirty years.}
\author{Jacques Raynal}
\date{  }
\maketitle
\centerline{\it  4 rue du Bief, 91380 Chilly--Mazarin, France}
\begin{abstract}
An expression for the spin--orbit interaction coupling between
different levels, which was shown to be aberrant more than thirty
years ago is used in a recent article \cite {RAP1} published in
Nuclear Physics. It leads to expressions quite simpler than they
should be.  Its behavior is in fact of a character opposite to that
of the spin--orbit interaction used in two--body studies.
\vskip .5truecm
PACS numbers 24.10.-i, 25.40.Dn, 25.40.Ny, 28.20.Cz
\end{abstract}

In the article entitled ``An algebraic solution of the multichannel
problem applied to low energy nucleon--nucleon scattering'', written by
K. Amos, L. Canton, G.Piseni, J. P. Svenne and D. van der Knijff,
published in Nucl. Phys. {\bf A728} (2003) 65--95 , the authors
describe in their Appendix B the interaction which they use.
On the second line of page 92~:
\begin{equation}
-{1\over {2\alpha r}} W_{ls} \{[\ell .\svec {s}]_{c'}
+[\ell .\svec {s}]_c\} \label{E1}
\end{equation}
with the two following lines deals with the spin--orbit
interaction. This is a generalisation of the
spin--orbit potential of the optical model, which is~:
\begin{equation}
{2\over r}\{{d\over dr}V(r)\}[\ell .\svec {s}]  \label{E2}
\end{equation}
and was used when first asymmetry measurements in inelastic
scattering with polarised proton beams became available.
The expression Eq. (\ref {E1}) was used in \cite{RAP2} without
being explicitly written in this article. It was sometimes
called Oak--Ridge term and gave results quite different from
the experimental results. Some attempts \cite {RAP3} were done
to get better results, adding $\ell ^ 2$ and $\svec {s}.\svec {I}$
terms which are also included in \cite{RAP1}.

Going back to the ``full Thomas term'' obtained for the spin--orbit
by transforming a Dirac equation into a Schr\"odinger equation,
J. S. Blair and H. Sherif \cite{RAP4,RAP5}, used the expression ~:
\begin{equation}
\nabla\{V(\svec{r})\} \times {\nabla \over i} .\svec{\sigma }
\label{E3}
\end{equation}
in computations for nucleon inelastic scattering and obtained a very
good reproduction of the experimental data.

This expression can be written \cite{RAP6,RAP7} like
Eq. (\ref {E1}). Using $V_{\lambda }(r)$ for the radial
dependence of a multipole $\lambda $, the result is~:
\begin{displaymath}
{1\over r}{d\over {dr}}V_{\lambda }(r)[\ell .\svec {s})_c
+{V_{\lambda }(r)\over {2r^2}}\bigg( {\lambda }({\lambda }+1)
-([\ell .\svec {s}]_{c'}-[\ell .\svec {s}]_c)
([\ell .\svec {s}]_{c'}-[\ell .\svec {s}]_c \pm 1)\bigg)
\end{displaymath}
\begin{equation}
+{V_{\lambda }(r)\over r}([\ell .\svec {s}]_c
-[\ell .\svec {s}]_{c'}){d\over {dr}}     \label{E4}
\end{equation}
with $+1$ or $-1$ if the wave function is not or is multiplied
by $r$. Note that there are here three form--factors~:
\begin{itemize}
\item ~~${1\over r}{d\over {dr}}V_{\lambda }(r)$
\hskip .5truecm
which is the only one for elastic scattering and is multiplied
only by the eigenvalue for the ket.
\item ~~${V_{\lambda }(r)\over {2r^2}}$
\hskip 1.2truecm
which is, divided by $r^2$, the true spin--orbit multipole which
does not appear in elastic scattering.
\item ~~${V_{\lambda }(r)\over r}{d\over {dr}}$
\hskip .7truecm
which is the form factor multiplying the derivative of the ket
radial function; integrating by part shows that the whole is
symmetric in $c$ and $c'$.
\end{itemize}

Except for the first, the form factors are not the ones of
Eq. (\ref {E1}). However, as the interaction of Eq. (\ref {E1})
is larger when the eigenvalues for $c$ and $c'$ are of the
same sign and the coefficients of the second and third
form--factors above are larger in the opposite case, one can
guess that the effect should be quite different.
The use of the spin--orbit deformation given by Eq. (\ref {E4})
in coupled channel calculation is not straightforward \cite{RAP8}.
It was the subject of codes {\bf \tt ECIS} (``{\it Equations
Coupl\'ees en It\'erations S\'equentielles}~'') from 
{\bf \tt ECIS68} to {\bf \tt ECIS97}. To compare results obtained
with the interactions given by Eq. (\ref {E1}) and Eq. (\ref {E3}),
the spin--orbit interaction is parametrised as~:
\begin{displaymath}
{1\over r}{d\over {dr}}V_{\lambda }(r)\bigg( z_1
+ z_3 [\ell .\svec {s})_c+ z_4 [\ell .\svec {s})_{c'}\bigg)
+{V_{\lambda }(r)\over r} z_6 ([\ell .\svec {s}]_c
-[\ell .\svec {s}]_{c'}){d\over {dr}}
\end{displaymath}
\begin{equation}
+{V_{\lambda }(r)\over {2r^2}} z_5 \bigg( { z_2 \lambda }({\lambda }+1)
-([\ell .\svec {s}]_{c'}-[\ell .\svec {s}]_c)
([\ell .\svec {s}]_{c'}-[\ell .\svec {s}]_c \pm 1)\bigg)     \label{E5}
\end{equation}
in all these codes. The coupling of Eq. (\ref {E1}) is obtained
by setting~:
\begin{equation}
z_1=z_2=z_5=z_6=0,  \qquad z_3=z_4={1\over 2},               \label{E6}
\end{equation}
the coupling of Eq. (\ref {E3}) multiplied by a parameter $\lambda $
(with $\lambda = 1$ for the ``unparametrised'' case) is given by~:
\begin{equation}
z_1=z_4=0, \qquad  z_2=1, \qquad  z_3=z_5=z_6=\lambda .      \label{E7}
\end{equation}
The parameter $\lambda $ allows to increase the strengh of the
spin--orbit transition without deforming its form factor in
the rotational model.
The use of $\lambda =2$ for experiments around 20 MeV gave
excellent results.

For people who don't want to consider Dirac equation at low
energy, there is another justification of Eq. (\ref {E3})
based on the nucleon--nucleon interaction \cite{RAP6,RAP7,RAP9}.
At the zero--range limit, the two--body spin--orbit interaction
of a nucleon exciting a particle--hole state is given by
Eq. (\ref {E4}) with the product of the particle and the hole
functions as $V_{\lambda }$, assuming that the sum of the
eigenvalues of $\ell .\svec {s}$ for the particles and holes vanishes
in the result. The overestimation by a factor 4 of the spin--orbit
interaction in the first publications on this subject
is an error which does not affect this similarity.
This approach, with the most general consideration of the
two--body interaction \cite{RBP1} has been the
subject of a series of codes, since {\bf \tt DWBA70} to
{\bf \tt DWBA98}. Note that spin--orbit interaction involves first
derivatives, quadratic spin--orbit involves second derivatives.
In this approach, the geometrical coefficients are expressed with
$\kappa =[\ell .\svec {s]}+1$ which is the coefficient of
${1\over r}$ in Dirac equation. For a central interaction
$V(r)(\svec{\sigma _1}.\svec{\sigma _2})$ and a tensor
interaction, the geometrical factor can be reduced to
the one involved by a spin--independent interaction leading to
two one--body form--factors~: one simple and one multiplied
by the sum or the difference of the $\kappa $. Usual and
quadratic spin--orbit need more form--factors. The two--body
interaction which keeps the parity for each body is characterised by
coefficient $\kappa _c - \kappa _{c'}$; the one which inverts
them is characterised by $\kappa _c + \kappa _{c'}$, equal to
a constant with what appears in Eq. (\ref {E1}) but has no
equivalent in one-body interaction (the spin--orbit has no
derivative but a third form--factor with
$(\kappa _c + \kappa _{c'})_2$).

The reference \cite{RBP1}, in which the first co--author is the
same as for \cite{RAP1} shows that some users of {\bf \tt DWBA}
or {\bf \tt ECIS} do not realise what these codes involve. In fact,
the author of this ``comment'' cannot answer on many points of
the article \cite{RBP1}.

There is no allusion to a second matrix with derivative
of the ket $|c>$ in \cite{RAP1}. In fact, its Eqs (36) and (42)
includes expressions like $[\ell .\svec{s}]_{c'c}$ which
could be a shorthand notation of a correct spin--orbit interaction;
this cannot convince the reader that the formulae of
Appendix B are not used. The model presented here is the
two-phonons vibrational model limited to one kind of phonons;
there is no factor $\kappa _c +\kappa _{c'}$ which appears only for two
phonons of angular momentum $L_1$ and $L_2$ coupled to a $J$
such that $L_1+L_2+J$ is odd.

The Eq. (\ref {E1}) is often written by authors to say that they
do not use it and cannot be found in \cite{RAP2,RAP3}. We cannot
know how many works used it since thirty years; the main interest
of \cite{RAP1} is to show that it continues to be used by some
physicists, at least from time to time. Its behaviour being opposite
to the one used in two--body studies and its lack of justification
(except the great simplification of the problem) justify the
qualification of ``aberrant''.

I thank the Service de Physique Th\'eorique de Saclay for allowing
me to follow this kind of problems after my retirement. I also thank
H. Sherif for reading the manuscript and for helpful correspondence.

\end{document}